\documentclass[pre,twocolumn,superscriptaddress,showpacs]{revtex4-2}
\usepackage{amsmath,amssymb,dsfont,graphicx}
\usepackage[utf8x]{inputenc}
\usepackage[T1]{fontenc}
\usepackage[english]{babel}

\newcommand*\rfrac[2]{{}^{#1}\!/_{#2}}
\newcommand\lp{\left(}
\newcommand\rp{\right)}
\newcommand\ls{\left[}
\newcommand\rs{\right]}
\newcommand\lbr{\left\lbrace}
\newcommand\rbr{\right\rbrace}

\begin{document}

\title{\mbox{Mean-field nature of synchronization stability in networks with multiple interaction layers}}
\date{\today}

\author{Charo I. del Genio}
\email{ad0364@coventry.ac.uk}
\affiliation{Centre for Fluid and Complex Systems, School of Computing, Electronics and Mathematics, Coventry University, Priory Street, Coventry CV1 5FB, UK}
\author{Sergio Faci-Lázaro}
\affiliation{Department of Condensed Matter Physics, University of Zaragoza, Zaragoza 50009, Spain}
\affiliation{GOTHAM lab, Institute of Biocomputation and Physics of Complex Systems (BIFI), University of Zaragoza, Zaragoza 50018, Spain}
\author{Jesús Gómez-Gardeñes}
\affiliation{Department of Condensed Matter Physics, University of Zaragoza, Zaragoza 50009, Spain}
\affiliation{GOTHAM lab, Institute of Biocomputation and Physics of Complex Systems (BIFI), University of Zaragoza, Zaragoza 50018, Spain}
\author{Stefano Boccaletti}
\affiliation{CNR-Institute of Complex Systems, Florence, Italy}
\affiliation{Moscow Institute of Physics and Technology, Dolgoprudny, Moscow Region, Russian Federation}
\affiliation{Universidad Rey Juan Carlos, Móstoles, Madrid, Spain}

\begin{abstract}
\begin{center}
 \textbf{ABSTRACT}
\end{center}
The interactions between the components of many real-world systems
are best modelled by networks with multiple layers. Different theories
have been proposed to explain how multilayered connections affect
the linear stability of synchronization in dynamical systems. However,
the resulting equations are computationally expensive, and therefore
difficult, if not impossible, to solve for large systems. To bridge
this gap, we develop a mean-field theory of synchronization for networks
with multiple interaction layers. By assuming quasi-identical
layers, we obtain accurate assessments of synchronization stability
that are comparable with the exact results. In fact, the accuracy of our
theory remains high even for networks with very dissimilar layers,
thus posing a general question about the mean-field nature of synchronization
stability in multilayer networks. Moreover, the computational complexity
of our approach is only quadratic in the number of nodes, thereby
allowing the study of systems whose investigation was thus far precluded.
\end{abstract}

\maketitle

\section*{Introduction}
The use of a network structure, consisting of a list of pairwise
connections called edges between discrete elements called nodes
or vertices, has long been a powerful abstraction to model and investigate
the behavior of complex systems~\cite{Alb02,New03,Boc06}. One area
in which the network paradigm has proved especially useful is the
study of distributed coupled dynamical systems~\cite{Bar08}, where
the nodes correspond to dynamical systems that interact across the
edges. Collective and organized dynamics is widely studied in such
networks, with a particular regard to the phenomenon of synchronization,
which holds fundamental importance in numerous natural
instances~\cite{Bar08,Gad95,Pec98,Lag00,Bar02,Boc02,Nis03,Bel04,Cha05,Hwa05,Mot05,Gom07,Yao09,Li13,Mas13,del15,Boc18}.

Within this topic, considerable attention has been paid to the stability
of synchronized states. A powerful method to estimate it is the so-called
Master Stability Function (MSF)~\cite{Pec98}, which allows one to compute
the value of the largest Lyapunov exponent $\Lambda_{\max}$ of the system
upon perturbation from the synchronized state. Then, one can determine whether
synchronization is stable simply from its sign: the perturbed system will
asymptotically re-synchronize only if $\Lambda_{\max}$ is negative. The
MSF has maintained its status as method of choice even when applied to
models that extend the network approach to the multilayer case~\cite{Boc14,Kiv14,Tan19}
and to networks with higher-order interactions~\cite{Gam21}, which better
capture the many levels of complexity of real-world systems.

In the dynamical multilayer networks we consider here,
the nodes are allowed to interact over multiple layers, each representing
a different type of interaction. Thus, they can be considered
a special case of the traditional multiplex networks, in which each
node is replicated in every layer, and interlayer links determine how
the state in each layer affects those of the neighbouring ones. Networks
with multiple interaction layers are often encountered in natural and
engineered complex systems of wide relevance. The paradigmatic example
is that of the nervous system, in which the same neurons are connected
to each other via two different types of synapses~\cite{Pur18}. Another
instance is provided by the transport infrastructure of large cities
where the same station is served by multiple means of mass transport.
In both cases, the value of a variable of interest, such as the potential
accumulated in a neuron or the state of congestion of a station, affects
equally all the processes involving the node~\cite{Sor12}.

The increase in complexity
of the system with respect to single-layer networks is
reflected by the fact that the use of the MSF no longer results
in a single equation, but rather produces a set of coupled linear
differential equations~\cite{del16}. Apart from its generality,
these master stability equations allowed researchers to identify
synchronized states whose stability properties are inherently due
to the multiplex architecture, such as stable coherent dynamics
in networked layers that are unstable when studied in isolation.
Also, the versatility of the MSF made it a method of
choice to study this type of networks under different conditions,
such as time-varying structures~\cite{Rak18}, as well as with the
help of other approaches, such as fast-switching techniques~\cite{Rak20}.
Unfortunately, despite the promising preliminary results, the master
stability equations also have a drawback, as the computational complexity
of solving the system makes it unwieldy for networks larger than
a couple of hundred nodes. This difficulty
is due to the mathematical structure that results from the MSF approach,
and therefore it does not depend on the specific network model studied or
on the particular formalism used upon it. In fact, even when the MSF takes
a fairly workable form, such as in traditional multiplex networks with dynamics
formalized via supra-Laplacians~\cite{Tan19}, the explicit numerical calculations
reach a high computational complexity if the supra-Laplacians do
not commute.

In this article, we introduce a mean-field theory of synchronization
for networks with multiple interaction layers. We derive our theory
in the assumption that the layers are quasi-identical, but
show that its range of applicability encompasses the case of very dissimilar
layers. In particular,
the estimate of the stability of synchronization obtained with our
theory virtually never changes with respect to the exact result, suggesting
that global synchronization stability in multilayer networks is inherently
a mean-field phenomenon. In addition, the numerical complexity of our
method is lower than that of the exact formulation,
making it applicable to large systems whose study would otherwise be
prevented by computational costs.

\section*{Results}
\subsection*{The model}
The derivation of a MSF on a networked system is effectively
a decomposition of the dynamics into eigenmodes. In a network
with $N$ nodes and $M$ interaction layers, the connection weights
of each layer $\alpha$ are the elements of the weighted adjacency
matrix $\textbf W^{(\alpha)}$. Let $\mathbf x_i$ be an $m$-dimensional
vector describing the state of node $i$, and $\textbf{F}: \mathbb R^m\rightarrow\mathbb R^m$
and $\textbf{H}_\alpha: \mathbb R^m\rightarrow\mathbb R^m$
be continuous and differentiable vector fields describing the
local dynamics and the interactions in layer $\alpha$, respectively.
Assuming diffusive coupling between the nodes, the global dynamics
is determined by the system
\begin{equation}\label{dynamics}
 \dot{\mathbf x}_i=\mathbf F\left(\mathbf{x}_i\right)-\sum_{\alpha=1}^M\sigma_\alpha\sum_{j=1}^NL_{i,j}^{\left(\alpha\right)}\mathbf H_{\alpha}\left(\mathbf{x}_j\right)\:.
\end{equation}
In Eqs.~\eqref{dynamics}, $\sigma_\alpha$ is the interaction strength
within layer $\alpha$, and $\mathbf L^{\left(\alpha\right)}$ is the graph
Laplacian of layer $\alpha$, whose diagonal elements are $L_{i,i}^{(\alpha)}=\sum_{j=1}^NW_{i,j}^{\left(\alpha\right)}$,
and whose off-diagonal elements are $L_{i,j}^{(\alpha)} = -W_{i,j}^{(\alpha)}$.

Following the approach by del~Genio \emph{et al.}~\cite{del16}, one can linearize Eqs.~\eqref{dynamics},
obtaining expressions for the evolution of the global synchronization error vector, which can be
projected onto the eigenvectors of the Laplacian of one of the layers. Choosing layer $\alpha=1$
(without loss of generality) results in the system
\begin{multline}\label{maineq}
 \dot{\boldsymbol\eta_j} = \lp J\mathbf F\lp\mathbf s\rp-\sigma_1\lambda_j^{(1)}J\mathbf H_1\lp\mathbf s\rp\rp\boldsymbol\eta_j+\\
  -\sum_{\alpha=2}^M\sigma_\alpha\sum_{k=2}^N\sum_{r=2}^N\lambda_r^{(\alpha)}\Gamma^{(\alpha)}_{r,k}\Gamma^{(\alpha)}_{r,j}J\mathbf H_{\boldsymbol\alpha}\lp\mathbf s\rp\boldsymbol\eta_k\:.
\end{multline}
Here, $\boldsymbol\eta$ is a vector of vectors whose $j$th component
is the projection of the global synchronization error vector onto the
space spanned by the $j$th Laplacian eigenvector of layer 1; $J$ is
the Jacobian operator; $\mathbf s$ is the state vector corresponding
to the synchronized state; $\lambda_j^{(\alpha)}$ is the $j$th Laplacian
eigenvalue of layer $\alpha$; and $\boldsymbol\Gamma^{(\alpha)}$ is
the spectral overlap matrix between layer 1 and layer $\alpha$, defined
as $\boldsymbol\Gamma^{(\alpha)}={\mathbf V^{(\alpha)}}^\mathrm T\mathbf V^{(1)}$,
where $\mathbf V^{(\alpha)}$ is the matrix of Laplacian eigenvectors
of layer $\alpha$ and T indicates transposition.

\subsection*{Mean-field theory}
To develop our theory, we start from the assumption
that the interaction layers are quasi-identical. Then,
due to their adjacency matrices and their Laplacians
being very similar, one can expect their Laplacian
eigenvectors to be equal up to some small perturbation.
Our goal is to spread the effect of this perturbation
in an equal way over all the directions transverse
to the synchronization manifold, which is identified
by the Laplacian eigenvector corresponding to the null
eigenvalue. To do so, compute the dynamical distance
between layer $\alpha$ and layer 1 \cite{del16}:
\begin{equation}\label{dyndist}
 \ell_{\alpha,1} = \sum_{i=2}^N\lbr\ls\sum_{j=2}^N\lp\Xi_{i,j}^{(\alpha,1)}\rp^2\rs-\lp\Xi_{i,i}^{(\alpha,1)}\rp^2\rbr\:.
\end{equation}
Note that, while here we consider dynamical
distances between any layer and the reference layer~1, in
principle this quantity is an indicator of the similarity
between the dynamics of any two layers. In fact, its general
definition is the sum of the squares of the off-diagonal
terms in the spectral overlap matrix between the two layers
considered, so that if the dynamics of the layers are identical
(i.e., their Laplacians commute), their dynamical distance
vanishes.
In the equation above, the matrices $\boldsymbol\Xi^{(\alpha,\beta)}$
are defined as $\boldsymbol\Xi^{(\alpha,\beta)}=\boldsymbol\Gamma^{(\alpha)}{\boldsymbol\Gamma^{(\beta)}}^{\mathrm T}={\mathbf V^{(\alpha)}}^{\mathrm T}\mathbf V^{(\beta)}$,
so that $\boldsymbol\Xi^{(\alpha,1)}={\mathbf V^{(\alpha)}}^{\mathrm T}\mathbf V^{(1)}=\boldsymbol\Gamma^{(\alpha)}$.
Also, the sums in Eq.~\eqref{dyndist} start from 2 because the first
Laplacian eigenvector is always $N^{-\rfrac{1}{2}}\lp1, \dotsc, 1\rp^\mathrm{T}$.
Therefore, $\Gamma^{(\alpha)}_{1,1}=1$ and $\Gamma^{(\alpha)}_{1,k}=\Gamma^{(\alpha)}_{k,1}=0$
for all $k>1$.
Now note that each $\boldsymbol\Gamma^{(\alpha)}$ is an
orthogonal matrix, since it is the product of two orthogonal
matrices. Moreover, from the definition, it follows
that $\mathbf V^{(\alpha)}=\mathbf V^{(1)}{\boldsymbol\Gamma^{(\alpha)}}^{\mathrm T}$.
In other words, $\boldsymbol\Gamma^{(\alpha)}$ is
the transformation matrix from the Laplacian eigenvectors
of the first layer to those of
layer $\alpha$. Then, our aim is replacing $\boldsymbol\Gamma^{(\alpha)}$
with a $\boldsymbol\Gamma^{(\alpha),\mathrm{MF}}$ whose action
is to change each eigenvector of the first layer
in the same fashion.

As $\boldsymbol\Gamma^{(\alpha)}$ is a rotation,
a natural choice is to make $\boldsymbol\Gamma^{(\alpha),\mathrm{MF}}$
rotate the eigenvectors of the first layer by the
same angle in every direction. More precisely, we
consider all possible 2-dimensional subspaces of
$\mathbb R^{N-1}$ determined by choosing any two
Laplacian eigenvectors of layer~1 except the first,
and then construct a matrix that rotates all the
Laplacian eigenvectors of layer~1 except the first
by the same amount in each of these subspaces. Note
that, in principle, this problem is underspecified,
as rotation matrices do not commute in 3 or more
dimensions, and the specific form of $\boldsymbol\Gamma^{(\alpha),\mathrm{MF}}$
depends on the order in which the rotations in the
individual subspaces are performed. However, since
the layers are quasi-identical, the rotation angle
needed is very small, as we will justify quantitatively
later on. Thus, the rotations that compose $\boldsymbol\Gamma^{(\alpha),\mathrm{MF}}$
are infinitesimal. In turn, this means that each
one of them, and indeed $\boldsymbol\Gamma^{(\alpha),\mathrm{MF}}$
itself, can be written as the sum of the identity
matrix and an element of the Lie algebra of $O\lp N-1\rp$,
or, more precisely, of $SO\lp N-1\rp$, since the
rotations are proper. Thus, for all $1<r<s$, the
elements of the matrix $\mathbf{R^{(r,s)}}$ that
operates the rotation in the subspace spanned by
the $r$th and $s$th eigenvectors are
\begin{equation*}
 R^{(r,s)}_{i,j} =
 \begin{cases}
  1 &\quad\text{if $i=j$}\\
  -\varepsilon^{(\alpha)} &\quad\text{if $i=r$ and $j=s$}\\
  \varepsilon^{(\alpha)} &\quad\text{if $i=s$ and $j=r$}\\
  0 &\quad\text{otherwise}\:,
 \end{cases}
\end{equation*}
where $\varepsilon^{(\alpha)}$ is the rotation angle, which
depends on the layer $\alpha$. Importantly, all the $\mathbf{R^{(r,s)}}$
commute, which removes the problem of considering the order
of the constitutive rotations of $\boldsymbol\Gamma^{(\alpha),\mathrm{MF}}$.
Taking into account all subspaces, and neglecting terms of
order higher than $\varepsilon^{(\alpha)}$, yields the following
form for the matrix:
\begin{equation}\label{GammaMF}
 \boldsymbol\Gamma^{(\alpha),\mathrm{MF}} = \begin{pmatrix}
                                      1 & 0                      & 0                       & 0                       & \hdots \\
                                      0 & 1                      & -\varepsilon^{(\alpha)} & -\varepsilon^{(\alpha)} & \hdots \\
                                      0 & \varepsilon^{(\alpha)} & 1                       & -\varepsilon^{(\alpha)} & \hdots \\
                                      0 & \varepsilon^{(\alpha)} & \varepsilon^{(\alpha)}  & 1                       & \ddots \\
                                      \vdots & \vdots & \vdots & \ddots & \ddots
                                     \end{pmatrix}\:.
\end{equation}
Notice that $\boldsymbol\Gamma^{(\alpha),\mathrm{MF}}$ is an orthogonal
matrix, as it should be, but only to first order, as expected from the
approximations used. Also note that the final results one obtains
using this formulation are the same that would be found using the exact
expressions on layers with a different structure. This mean-field-equivalent
structures can be found by first computing the mean-field-equivalent Laplacian
eigenvectors using the same equation as above, namely
$\mathbf V^{(\alpha),\mathrm{MF}}=\mathbf V^{(1)}\lp\boldsymbol\Gamma^{(\alpha),\mathrm{MF}}\rp^{\mathrm T}$,
and then using them to recover the mean-field-equivalent Laplacian itself.
To find the value of $\varepsilon^{(\alpha)}$ to be used in Eq.~\eqref{GammaMF},
one can notice that
$\sqrt{\sum_{i=2}^N\lbr\ls\sum_{j=2}^N\lp\boldsymbol\Gamma^{(\alpha),\mathrm{MF}}_{i,j}\rp^2\rs-\lp\boldsymbol\Gamma^{(\alpha),\mathrm{MF}}_{i,i}\rp^2\rbr} = \sqrt{\lp N-1\rp\lp N-2\rp{\varepsilon^{(\alpha)}}^2}=\varepsilon^{(\alpha)}\sqrt{\lp N-1\rp\lp N-2\rp}$.
But from Eq.~\eqref{dyndist} and the fact
that $\boldsymbol\Xi^{(\alpha,1)}=\boldsymbol\Gamma^{(\alpha)}$
it follows that this quantity has to be
equal to $\sqrt{\ell_{\alpha,1}}$. Thus,
\begin{equation}\label{epsi}
 \varepsilon^{(\alpha)} = \sqrt{\frac{\ell_{\alpha,1}}{\lp N-1\rp\lp N-2\rp}}\:.
\end{equation}

Note that Eq.~\eqref{epsi} effectively
provides a mean of the dynamical distance over all
the directions transverse to the synchronization manifold,
thereby justifying our definition of the method as
a mean-field theory. Finally, to find the mean-field form of the MSF,
one can notice that Eq.~\eqref{maineq} contains the product
$\Gamma^{(\alpha)}_{r,k}\Gamma^{(\alpha)}_{r,j}$
within its innermost sum. Then, replacing the matrices
$\boldsymbol\Gamma^{(\alpha)}$ with their mean-field
version, one obtains
\begin{equation*}
 \Gamma^{(\alpha),\mathrm{MF}}_{i,j} =
 \begin{cases}
  1 &\quad\text{if $i=j$}\\
  -\varepsilon^{(\alpha)} &\quad\text{if $j>i$ and $i>1$}\\
  \varepsilon^{(\alpha)} &\quad\text{if $j<i$ and $j>1$}\\
  0 &\quad\text{otherwise}\:.
 \end{cases}
\end{equation*}
Since, in Eq.~\eqref{maineq}, $j$, $r$ and $k$ are always greater
than 1, one can write
\begin{equation*} \label{pippino}
 \Gamma^{(\alpha),\mathrm{MF}}_{r,k}\Gamma^{(\alpha),\mathrm{MF}}_{r,j} =
 \begin{cases}
  1 & \quad\text{if $r=k=j$}\\
  -\varepsilon^{(\alpha)} & \quad\substack{\text{if\ }r=k\ \text{and\ }j>r,\\\text{or if\ }k>r\ \text{and\ }r=j}\\
  \varepsilon^{(\alpha)} & \quad\substack{\text{if\ }r=k\ \text{and\ }j<r,\\\text{or if\ }k<r\ \text{and\ }r=j}\\
  {\varepsilon^{(\alpha)}}^2\approx 0 & \quad\substack{\text{if\ }k>r\ \text{and\ }j>r,\\\text{or if\ }k<r\ \text{and\ }j<r}\\
  -{\varepsilon^{(\alpha)}}^2\approx 0 & \quad\substack{\text{if\ }k>r\ \text{and\ }j<r,\\\text{or if\ }k<r\ \text{and\ }j>r}\:.
 \end{cases}
\end{equation*}
Using the expression above, the second term on the right-hand
side of Eq.~\eqref{maineq} becomes
\begin{multline}\label{MFMSF}
  \sum_{\alpha=2}^M\sigma_\alpha\Bigg[\lambda_j J\mathbf H_{\boldsymbol\alpha}\lp\mathbf s\rp\boldsymbol\eta_j+\\
  \sum_{k=2}^{j-1} \varepsilon^{(\alpha)}\lp\lambda_j^{(\alpha)}-\lambda_k^{(\alpha)}\rp J\mathbf H_{\boldsymbol\alpha}\lp\mathbf s\rp\boldsymbol\eta_k+\\
  \sum_{k=j+1}^N \varepsilon^{(\alpha)}\lp\lambda_k^{(\alpha)}-\lambda_j^{(\alpha)}\rp J\mathbf H_{\boldsymbol\alpha}\lp\mathbf s\rp\boldsymbol\eta_k\Bigg]\:,
\end{multline}
where we have split the sum over $k$ into $k=j$,
$k<j$ and $k>j$.

The expression~\eqref{MFMSF} greatly simplifies the calculations
with respect to the original formulation. In fact, the computational
complexity of calculating each component of $\boldsymbol\eta$ according
to Eq.~\eqref{maineq} is $O\lp N^2M\rp$, whereas using Eq.~\eqref{MFMSF},
this reduces to $O\lp NM\rp$. Also note that in Eq.~\eqref{MFMSF},
the first term inside the sum over $\alpha$ corresponds to the case
of commuting Laplacians. Thus, while the transverse
modes are still not completely decoupled, which would only be possible
in the commuting case, our theory effectively consists in a first-order
correction, obtained by a mean-field
perturbative approximation of the dynamics. This becomes even more
evident when rewriting the whole Eq.~\eqref{maineq} as
\begin{multline}\label{MFsync}
 \dot{\boldsymbol\eta_j} = \lp J\mathbf F\lp\mathbf s\rp-\sum_{\alpha=1}^M\sigma_\alpha\lambda_j^{(\alpha)} J\mathbf H_{\boldsymbol\alpha}\lp\mathbf s\rp\rp\boldsymbol\eta_j-\\
 \sum_{\alpha=2}^M \varepsilon^{(\alpha)}\sigma_\alpha\ls\sum_{k=2}^N\left|\lambda_j^{(\alpha)}-\lambda_{k}^{(\alpha)}\right| J\mathbf H_{\boldsymbol\alpha}\lp\mathbf s\rp\boldsymbol\eta_k\rs\:,
\end{multline}
for which we assume that the Laplacian eigenvalues
are sorted in a non-decreasing way.
\begin{figure}[t!]
\includegraphics[width=0.45\textwidth]{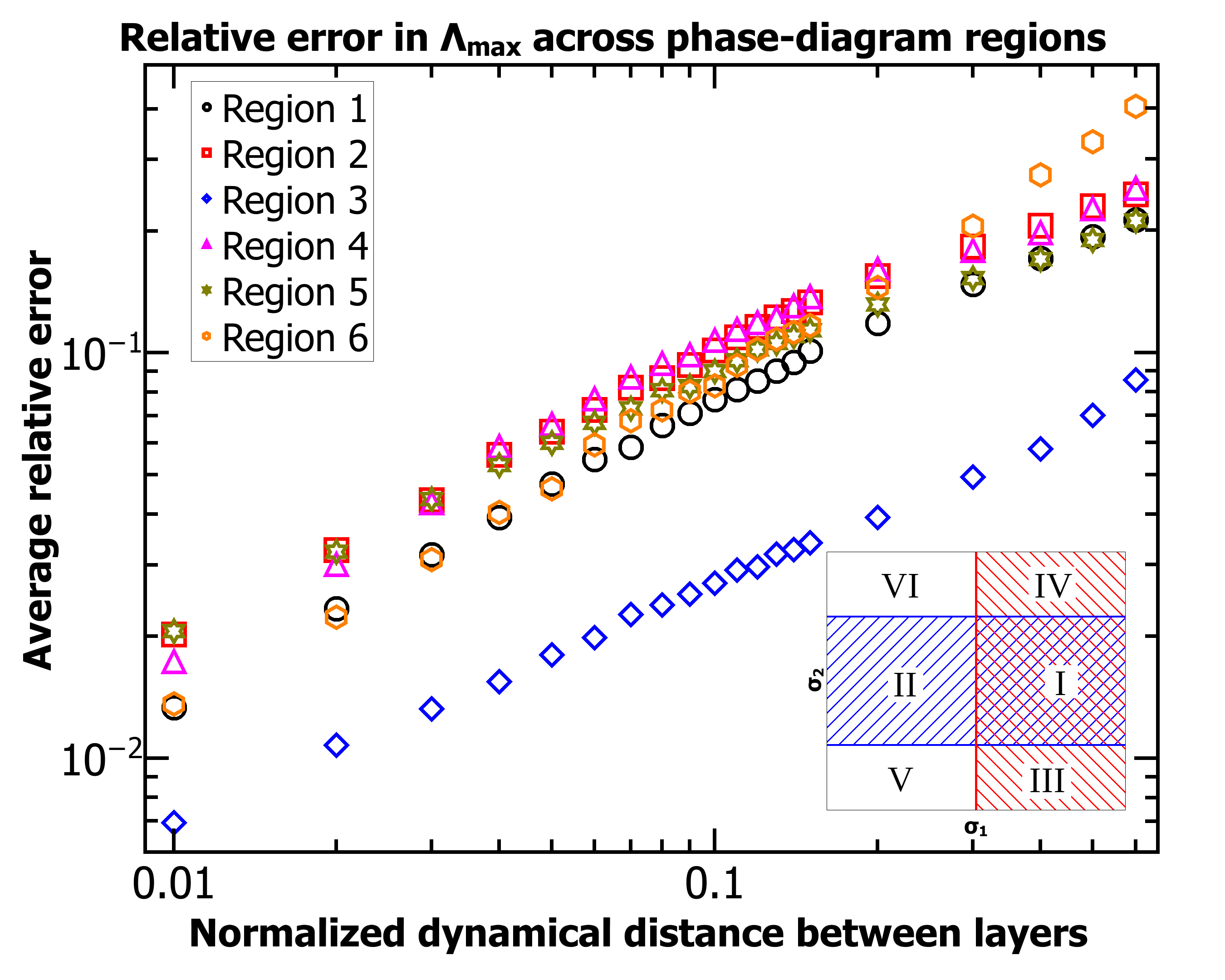}
\caption{{\bf The accuracy of our theory increases with the ratio of the interaction strengths
of the first layer $\sigma_1$ to that of the second layer $\sigma_2$}.
The relative error in the estimate of the largest Lyapunov exponent $\Lambda_{\max}$ of the perturbed system
decreases from a maximum of approximately $40\%$ in region~6 ($\rfrac{\sigma_1}{\sigma_2}\approx 0.0917$)
to a minimum of approximately $0.7\%$ in region~3 ($\rfrac{\sigma_1}{\sigma_2}\approx 3.5$).
Within each region, the error increases sublinearly with the normalized dynamical distance.
Each point is averaged over 1,000 realizations; error bars are smaller than the symbol size.
Inset: schematic illustration of the six regions (adapted from del~Genio \emph{et al.}~\cite{del16}). Layer~1 is
individually stable only when $\sigma_1$ is greater than a critical value (red striped regions);
layer~2 is individually stable only when $\sigma_2$ lies between two critical values (blue striped
regions); region 1 is the only zone of the phase diagram where both layers are already individually stable.}
\label{RegRelErr}
\end{figure}

The derivation of Eq.~\eqref{MFsync}, which constitutes
the mean-field approximation of Eq.~\eqref{maineq}, not
only offers a decrease in the complexity of
evaluating the linear stability of this type
of multiplexes, but also,
more importantly, paves the way to compute the stability
diagram of systems whose size is too large to permit the
use of Eq.~\eqref{maineq}. For example, a scaling
test reveals that simulation of a single duplex random
network with 10000 nodes would take approximately 103~years
on a workstation with a 16-core Intel\textregistered\ Xeon\textregistered\ Gold
6130 CPU if using Eq.~\eqref{maineq}. This time would
be reduced to just above 5~days if using Eq.~\eqref{MFMSF}.
\begin{figure}[t!]
\includegraphics[width=0.45\textwidth]{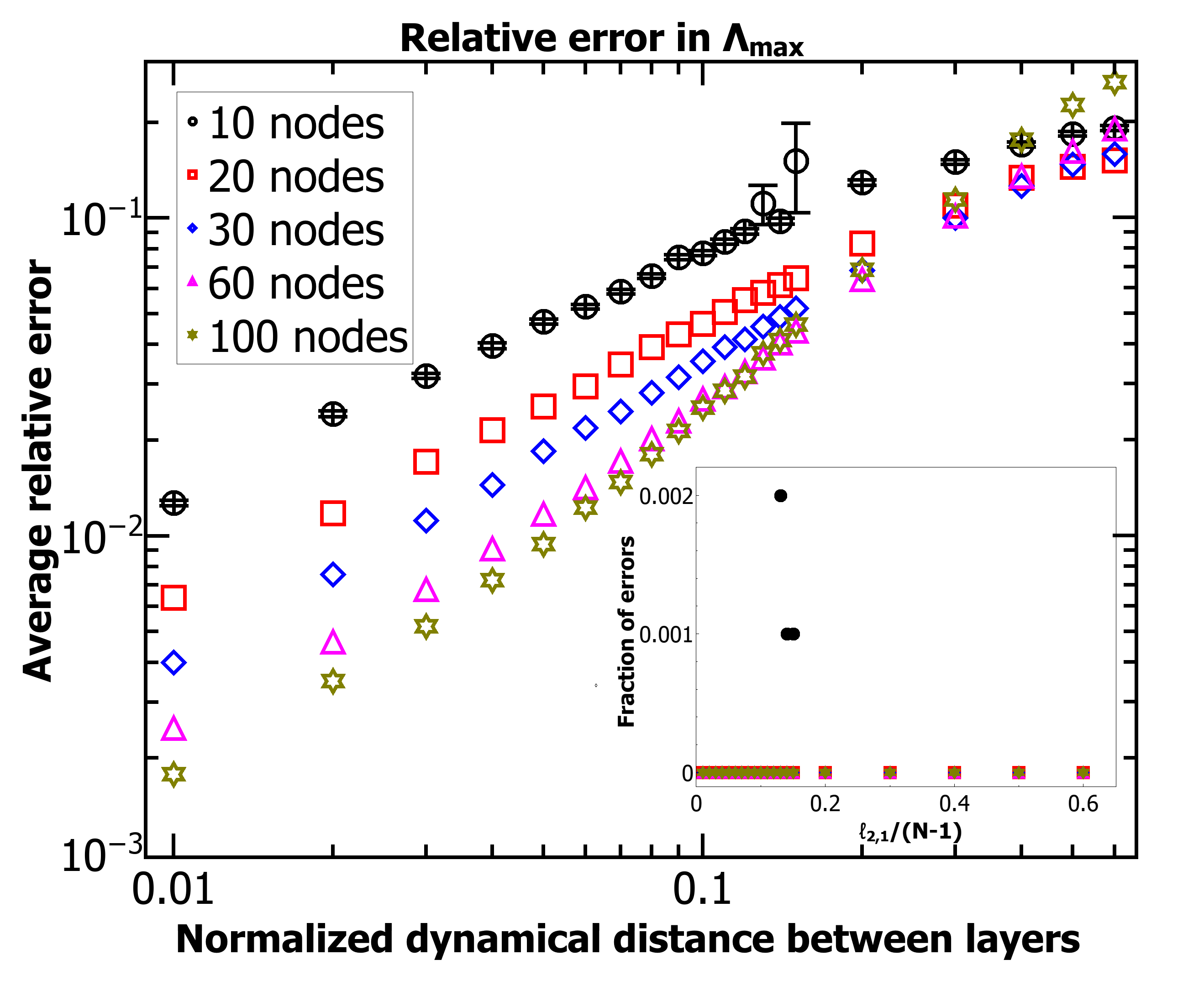}
\caption{{\bf Synchronization stability of networks
with multiple interaction layers is virtually always mean-field}.
In networks with $N$ nodes per layer, for normalized
dynamical distances $\ell_{2,1}/(N-1)$ smaller than $0.1$, the accuracy
of the estimate of the maximum Lyapunov exponent $\Lambda_{\max}$ improves with
system size. Also, the error in the estimate never
increases beyond $0.3$, even for very large dynamical
distances. Each point is averaged over 1,000 realizations;
when not drawn, error bars are smaller than the symbol
size. Inset: our mean-field theory assesses synchronization
stability always correctly, with possible occasional
exceptions only occurring in very small systems.}
\label{RelErr}
\end{figure}

\subsection*{Numerical validation}
To demonstrate the validity and applicability
of our theory, we carried
out extensive numerical simulations on
2-layer random networks of chaotic Rössler oscillators.
Note that the dynamical distance
between layers 2 and 1, $\ell_{2,1}$,
is bound between 0 and $N-1$. Thus, to
allow for a direct comparison between
systems of different size, we henceforth
plot our results as a function of the
normalized dynamical distance $\tilde{\ell}_{2,1}=\rfrac{\ell_{2,1}}{\lp N-1\rp}$.
In Fig.~\ref{RegRelErr} we show the average error
in the determination of the largest Lyapunov exponent
$\Lambda_{max}$ as a function of $\tilde{\ell}$.
Our results indicate that the accuracy of the theory
is generally high, but with a marked dependence on
the region of the layer-stability phase diagram. More specifically,
the accuracy increases with the ratio $\rfrac{\sigma_1}{\sigma_2}$,
which is on average different between the regions
of the phase diagram (schematically illustrated
in the inset of Fig.~\ref{RegRelErr}).
This is consistent with our theory being a perturbative
correction to the commuting case, since the relative
contribution of such a correction is indeed inversely
proportional to the ratio of the interaction strengths
within the layers. At the same time, and as we stressed
above, the choice of the reference layer~1 is entirely
arbitrary. As such, one can always choose the layer
with the largest interaction strength to be layer~1,
which implies that the actual worst case scenario
is the one in which all layers have the same interaction
strength.

At the light of these considerations, we performed additional
simulations, now imposing $\sigma_1=\sigma_2$, to evaluate the
performance of our theory for increasing system sizes. The results,
illustrated in Fig.~\ref{RelErr}, show that the theory provides
accurate results for a wide range of normalized dynamical distances
between layers.

To further explore the scope of applicability
of our method, we carried out extra simulations on networks
consisting of three random layers, as well as on preferential-attachment
networks~\cite{Bar99}. The results are presented
in three figures (Supplementary Figures~1--3) of the Supplementary Information.
The case of 3-layer networks shows a general
improvement with respect to the 2-layer case, which is particularly
evident for larger systems sizes (see Supplementary Figure~1). Also, the errors
in assessment of synchronization stability completely vanish,
even for smaller networks. The situation is slightly different
for preferential-attachment networks, where the errors in
assessing stability are not identically null, but they remain
fairly negligible, reaching a maximum of $0.8\%$ at normalized
dynamical distances of $0.3$ or more (see Supplementary Figure~2). The average
relative errors in estimating the maximum Lyapunov exponent
remain instead roughly in line with those for random networks,
but without a clear trend with network size (see Supplementary Figure~3).

\section*{Discussion}
The simulation results illustrated above show that the error in the estimation of $\Lambda_{\max}$
is never larger than $30\%$, even for normalized dynamical distances
as high as $\tilde{\ell}=0.6$. Importantly, when $\tilde{\ell}<0.1$,
the accuracy of the estimate increases with the system size.
This is most likely a direct consequence of Eq.~\eqref{epsi},
which shows that, for the same value of the normalized dynamical
distance, $\varepsilon^{(\alpha)}$ decreases proportionally to
$\sqrt{N}$.

The applicability of our theory increases even more when considering
the typical use of the MSF, namely the assessment of the stability of
synchronization. In this case, our method virtually never fails in correctly
identifying stable and unstable synchronized states. In our simulations,
we only found occasional errors for at most $0.2\%$ of the cases, and
exclusively in networks of only 10~nodes. Given the very small size
of such systems, we believe these errors to have been caused by finite-size
effects, indicating that our theory always allows a correct determination
of synchronization stability, and it holds true far beyond infinitesimal
dynamical distances.

Note that, as this method is an approximation,
we do expect that it will present limitations. At the current stage,
we can only speculate that, in a manner similar to other mean-field
theories, it may stop providing correct results in the vicinity of critical
transitions, or for networks with pathological structures. Such cases
are however beyond the scope of the present work, and they will be
explored in future studies.

In conclusion, we developed a mean-field theory of synchronization
stability for networks with multiple interaction layers.
In principle, the same approach could be applied to different types
of networks, most likely resulting in similar but different equations.
While the theory has been derived under the assumption of quasi-identical
layers, we have shown that its range of validity and applicability
includes the case of very different layers.
In fact, our theory provides an accurate
assessment of synchronization stability in networks whose layers are
actually substantially different from each other. Moreover, the accuracy
of the predictions increases with system size, raising the question
of whether, in the thermodynamic limit, the linear stability of the
globally synchronized state becomes a pure mean-field effect. In addition, the numerical
complexity of our approach is lower than that of the
exact solution of the problem. These considerations make the application
of our theory particularly attractive in the case of large natural
systems, whose studies have been so far frustrated, if not completely
inhibited, and which may now become tractable both analytically and
computationally.

\section*{Methods}
\subsection*{Network structures}
The simulations performed on Erdős-Rényi used
a probability of occurrence of edges $p=0.4$.
This choice was made to guarantee a high probability
that the layers were connected, which was tested
for in each case, and which is a necessary condition
for the existence of a globally synchronized
state. Simulations on layers with
heterogeneous topology were carried out on preferential-attachment
layers with the same density.
After creating the reference layer, the second layer
and third layer were obtained
by perturbing the edges of the first using a
doubling-bisecting scheme to target the desired
normalized dynamical distance with a tolerance
of $10^{-4}$.

\subsection*{Layer dynamics}
The local dynamics
of Rössler oscillators is described by
$\mathbf x = (x_1,x_2,x_3),\ \mathbf{F}\lp\mathbf x\rp=(-x_2-x_3,x_1+ax_2,b+(x_1-c)x_3)^{\mathrm T}$.
Here, we chose $a=b=0.2$ and
$c=9$ to ensure chaotic local dynamics.
We also let the interaction functions for the layers
be $\mathbf{H_1}\lp\mathbf x\rp=(0,x_2,0)$
and $\mathbf{H_2}\lp\mathbf x\rp=(x_1,0,0)$,
because these choices are known to create a rich phase
diagram with 6 distinct regions of behaviour
determined by the combinations of synchronization
stability of the individual layers~\cite{del16}.
The third layer in the 3-layer
simulations was given interaction function
$\mathbf{H_3}\lp\mathbf x\rp=(x_1,0,0)$.

\subsection*{ODE integration}
The differential equation systems were integrated
using a Runge-Kutta-Fehlberg 4(5) method,
with step size $0.01$ and tolerance $10^{-6}$.
The value of each initial component
of the projected global synchronization error vector
transverse to the synchronization manifold
was chosen independently from a Gaussian distribution.
This guarantees that, upon normalization,
the initial projected transverse synchronization
error vector was a uniform random unit vector,
due to the spherical symmetry of multivariate
normal distributions.

\subsection*{Estimate of maximum Lyapunov exponents}
To compute the maximum Lyapunov exponents,
after a transient time of 50, we evolved the
systems for 500 windows of 100 integration
steps each. After each window, we computed
the logarithm of the norm of the components
of the projected global synchronization errors
transverse to the synchronization manifold,
and normalized them back to a unit vector.
Their averages provided estimates for the
maximum Lyapunov exponents sought.
${}$\\

\section*{Author contributions}
CIDG and SB developed the theory. CIDG, SFL and JGG performed
the simulations and analyzed the results. All authors wrote the
manuscript.

\section*{Data availability}
All data generated is present within the manuscript.

\begin{acknowledgments}
CIDG acknowledges support from UKRI under Future Leaders Fellowship
grant number MR/T020652/1. SFL and JGG acknowledge support from MINECO
and FEDER funds (Projects No. FIS2017-87519-P, No. FIS2017-90782-REDT
(IBERSINC), and No. PID2020-113582GB-I00) and from the Departamento
de Industria e Innovación del Gobierno de Aragón y Fondo Social Europeo
(FENOL group, grant E36-20r). SFL acknowledges financial support by Gobierno
de Aragón through the Grant defined in ORDEN IIU/1408/2018.
\end{acknowledgments}


\begin{thebibliography}{99}
\item[]\begin{center}\textbf{REFERENCES}\\${}$\\\end{center}
\bibitem{Alb02} Albert, R. \& Barabási, A.-L. Statistical mechanics of complex networks. \textit{Rev.\ Mod.\ Phys.} \textbf{74}, 47--97 (2002).
\bibitem{New03} Newman, M.~E.~J. The structure and function of complex networks. \textit{SIAM Review} \textbf{45}, 167--256 (2003).
\bibitem{Boc06} Boccaletti, S., Latora, V., Moreno, Y., Chavez, M. \& Hwang, D.-U. Complex networks: Structure and dynamics. \textit{Phys.\ Rep.} \textbf{424}, 175--308 (2006).
\bibitem{Bar08} Barrat, A., Barthélemy, M. \& Vespignani, A. \textit{Dynamical Processes on Complex Networks} (Cambridge University Press, 2008).
\bibitem{Gad95} Gade, P.~M., Cerdeira, H.~A. \& Ramaswamy, R. Coupled maps on trees. \textit{Phys.\ Rev.\ E} \textbf{52}, 2478--2485 (1995).
\bibitem{Pec98} Pecora, L.~M. \& Carroll, T.~L. Master Stability Functions for Synchronized Coupled Systems. \textit{Phys.\ Rev.\ Lett.} \textbf{80}, 2109--2112 (1998).
\bibitem{Lag00} Lago-Fernández, L.~F., Huerta, R., Corbacho F. \& Sigüenza, J.~A. Fast Response and Temporal Coherent Oscillations in Small-World Networks. \textit{Phys.\ Rev.\ Lett.} \textbf{84}, 2758 (2000).
\bibitem{Bar02} Barahona M. \& Pecora, L.~M. Synchronization in Small-World Systems. \textit{Phys.\ Rev.\ Lett.} \textbf{89}, 054101 (2002).
\bibitem{Boc02} Boccaletti, S., Kurths, J., Osipov, G., Valladares, D.~L. \& Zhou, C.~S., The synchronization of chaotic systems. \textit{Phys.\ Rep.} \textbf{366}, 1--101 (2002).
\bibitem{Nis03} Nishikawa, T., Motter, A.~E., Lai, Y.-C. \& Hoppensteadt, F.~C. Heterogeneity in Oscillator Networks: Are Smaller Worlds Easier to Synchronize? \textit{Phys.\ Rev.\ Lett.} \textbf{91}, 014101 (2003).
\bibitem{Bel04} Belykh, I.~V., Belykh, V.~N. \& Hasler, M. Blinking model and synchronization in small-world networks with a time-varying coupling. \textit{Physica~D} \textbf{195}, 188--206 (2004).
\bibitem{Cha05} Chavez, M., Hwang, D.-U., Amann, A., Hentschel, H.~G.~E. \& Boccaletti, S. Synchronization is Enhanced in Weighted Complex Networks. \textit{Phys.\ Rev.\ Lett.} \textbf{94}, 218701 (2005).
\bibitem{Hwa05} Hwang, D.-U., Chavez, M., Amann A. \& Boccaletti, S. Synchronization in Complex Networks with Age Ordering. \textit{Phys.\ Rev.\ Lett.} \textbf{94}, 138701 (2005).
\bibitem{Mot05} Motter, A.~E., Zhou C.~S. \& Kurths, J. Enhancing complex-network synchronization. Europhysics Letters \textbf{69}, 334--340 (2005).
\bibitem{Gom07} Gómez-Gardeñes, J., Moreno, Y. \& Arenas, A. Paths to Synchronization on Complex Networks. \textit{Phys.\ Rev.\ Lett.} \textbf{98}, 034101 (2007).
\bibitem{Yao09} Yao, J., Wang, H.~O., Guan Z.-H. \& Xu, W. Passive stability and synchronization of complex spatio-temporal switching networks with time delays. \textit{Automatica} \textbf{45}, 1721--1728 (2009).
\bibitem{Li13}  Li F. \& Lu, X. Complete synchronization of temporal Boolean networks. \textit{Neural Netw.} \textbf{44}, 72--77 (2013).
\bibitem{Mas13} Masuda, N., Klemm K. \& Eguíluz, V.~M. Temporal Networks: Slowing Down Diffusion by Long Lasting Interactions. \textit{Phys.\ Rev.\ Lett.} \textbf{111}, 188701 (2013).
\bibitem{del15} del~Genio, C.~I., Romance, M., Criado, R. \& Boccaletti, S. Synchronization in dynamical networks with unconstrained structure switching. \textit{Phys.\ Rev.~E} \textbf{92}, 062819 (2015).
\bibitem{Boc18} Boccaletti, S., Pisarchik, A.~N., del~Genio, C.~I. \& Amann, A. \textit{Synchronization: from coupled systems to complex networks} (Cambridge University Press, 2018).
\bibitem{Boc14} Boccaletti, S. \emph{et al.} The structure and dynamics of multilayer networks. \textit{Phys.\ Rep.} \textbf{544}, 1--122 (2014).
\bibitem{Kiv14} Kivela, M. \emph{et al.} Multilayer networks. \textit{J. Complex Networks} \textbf2, 203--271 (2014).
\bibitem{Tan19} Tang L., Wu, X., Lü, J., Lu, J. \& D'Souza, R.~M. Master stability functions for complete, intralayer, and interlayer synchronization in multiplex networks of coupled Rössler oscillators. \textit{Phys.\ Rev.~E} \textbf{99}, 012304 (2019).
\bibitem{Gam21} Gambuzza, L.~V. \emph{et al.} Stability of synchronization in simplicial complexes. \textit{Nat.\ Commun.} \textbf{12}, 1255 (2021).
\bibitem{Pur18} Purves, D. \emph{et al.} (eds.), \textit{Neuroscience} (Oxford University Press, 2018).
\bibitem{Sor12} Sorrentino, F. Synchronization of hypernetworks of coupled dynamical systems. \textit{New J.\ Phys.} \textbf{14}, 033035 (2012).
\bibitem{del16} del~Genio, C.~I., Gómez-Gardeñes, J., Bonamassa, I. \& Boccaletti, S. Synchronization in networks with multiple interaction layers. \textit{Sci.\ Adv.} \textbf{6}, e1601679 (2016).
\bibitem{Rak18} Rakshit, S., Bera, B.~K., Bollt, E.~M. \& Ghosh, D. Intralayer Synchronization in Evolving Multiplex Hypernetworks: Analytical Approach. \textit{SIAM J. Appl.\ Dyn.\ Syst.} \textbf{19}, 918--963 (2018).
\bibitem{Rak20} Rakshit, S., Bera, B.~K. \& Ghosh, D. Synchronization in a temporal multiplex neuronal hypernetwork. \textit{Phys.\ Rev.~E} \textbf{98}, 032305 (2020).
\bibitem{Bar99} Barabási, A.-L. \& Albert, R. Emergence of Scaling in Random Networks. \textit{Science} \textbf{286}, 509--512 (1999).
\end{thebibliography}
\end{document}


\title{Mean-field nature of synchronization stability in networks\\with multiple interaction layers -- Supplementary Information}
\date{\today}

\author{Charo I. del Genio}
\affiliation{Centre for Fluid and Complex Systems, School of Computing, Electronics and Mathematics, Coventry University, Priory Street, Coventry CV1 5FB, UK}
\author{Sergio Faci-Lázaro}
\affiliation{Department of Condensed Matter Physics, University of Zaragoza, Zaragoza 50009, Spain}
\affiliation{GOTHAM lab, Institute of Biocomputation and Physics of Complex Systems (BIFI), University of Zaragoza, Zaragoza 50018, Spain}
\author{Jesús Gómez-Gardeñes}
\affiliation{Department of Condensed Matter Physics, University of Zaragoza, Zaragoza 50009, Spain}
\affiliation{GOTHAM lab, Institute of Biocomputation and Physics of Complex Systems (BIFI), University of Zaragoza, Zaragoza 50018, Spain}
\author{Stefano Boccaletti}
\affiliation{CNR-Institute of Complex Systems, Florence, Italy}
\affiliation{Moscow Institute of Physics and Technology, Dolgoprudny, Moscow Region, Russian Federation}
\affiliation{Universidad Rey Juan Carlos, Móstoles, Madrid, Spain}
\maketitle

\begin{figure}[h]
\includegraphics[width=0.65\textwidth]{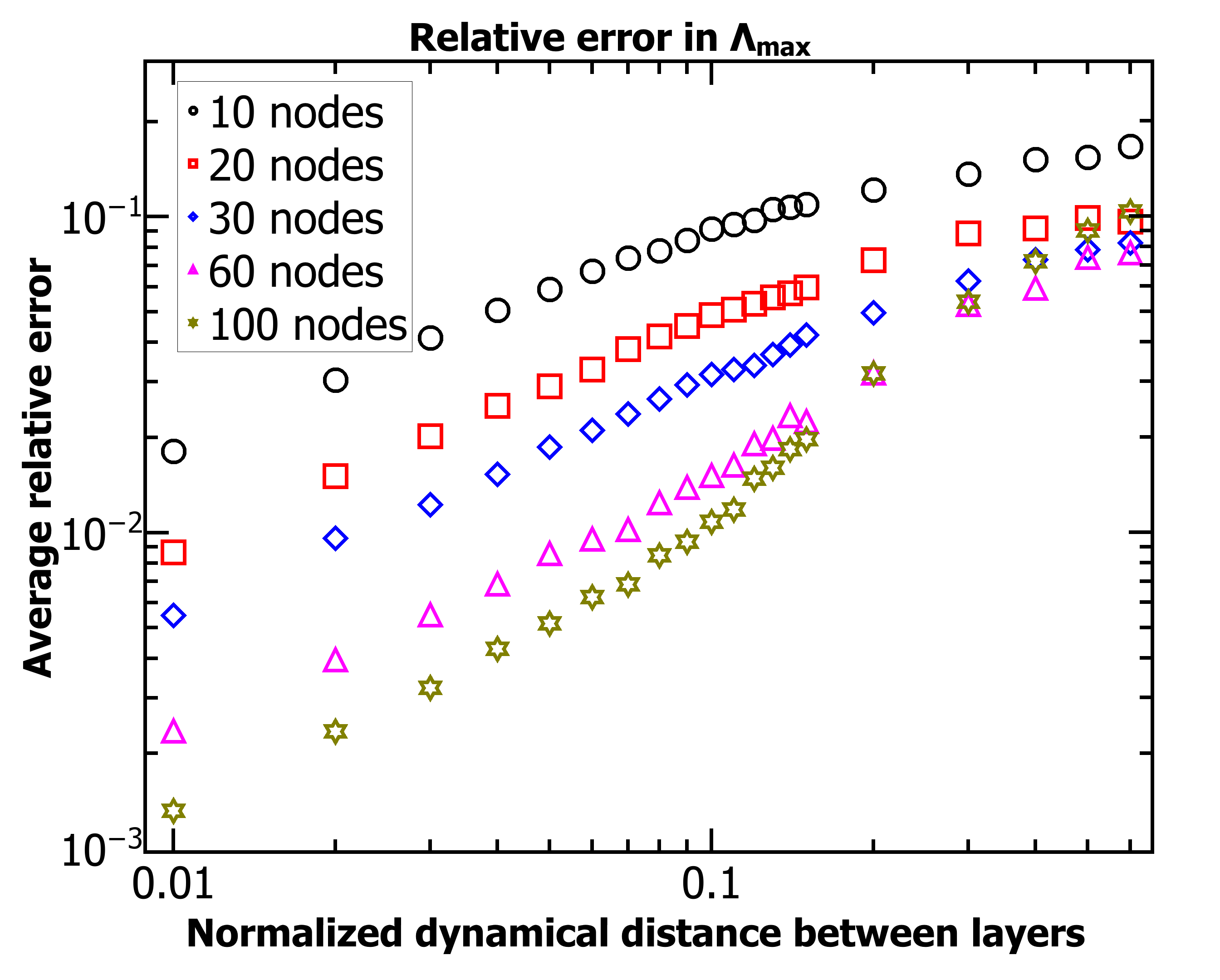}
\caption{{\bf The relative error in the largest Lyapunov exponent $\Lambda_{\max}$ decreases with the number of layers}.
The relative error in the estimate of the largest Lyapunov exponent of the perturbed system
in 3-layer networks is generally lower than that for 2-layer networks. This effect is particularly
evident for the largest system size, consistently with the scaling considerations about the rotation angle $\varepsilon^{(\alpha)}$.
Also, the range over which the accuracy increases with system size is larger than it is for 2-layer
networks. Each point is averaged over 1,000 realizations; error bars are smaller than the symbol size.}
\end{figure}

\begin{figure}[h]
\includegraphics[width=0.65\textwidth]{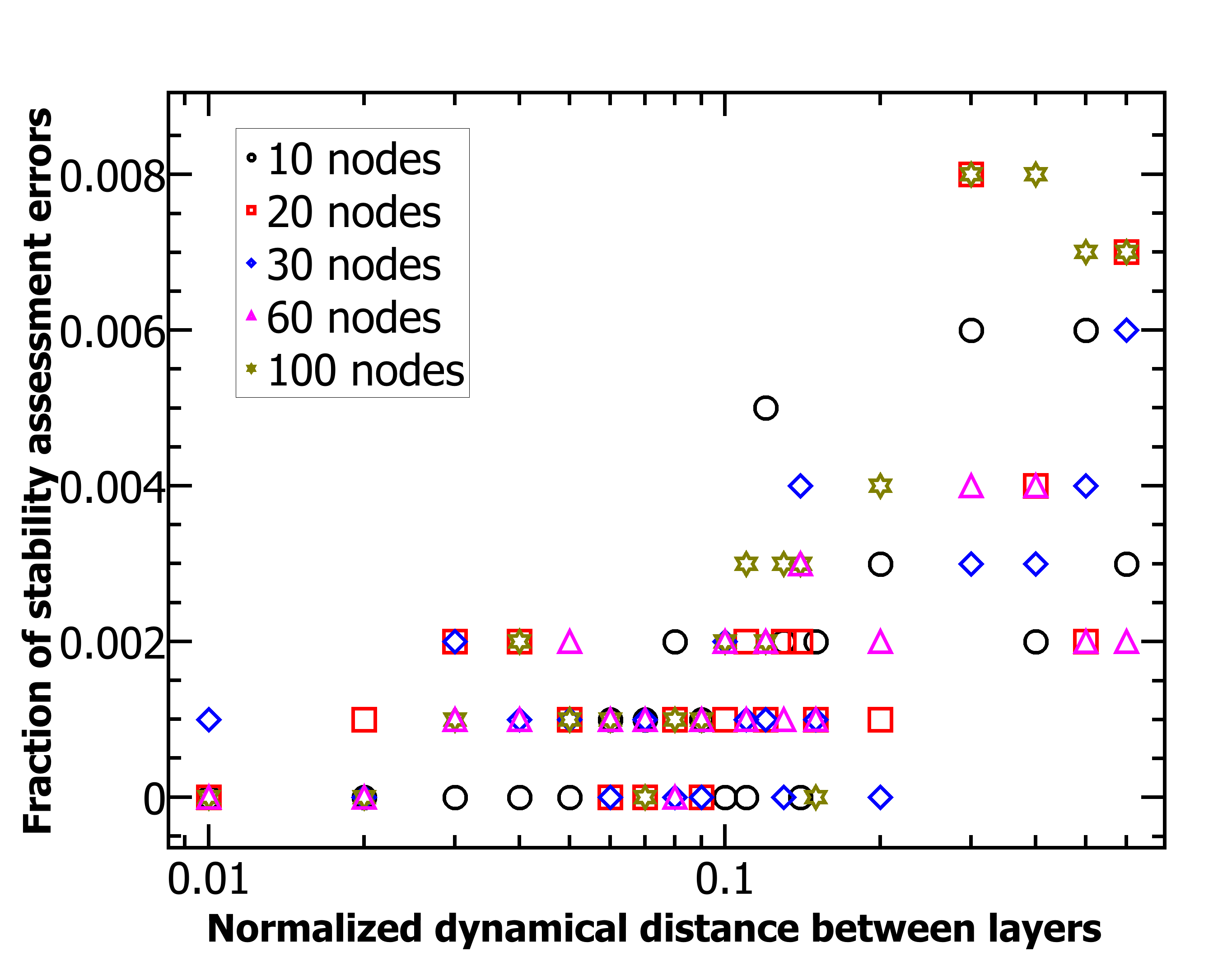}
\caption{{\bf Fraction of stability assessment errors
in preferential-attachment networks}.
In networks with heterogeneous layer topology,
the errors in assessing the synchronization stability
no longer vanish. Nonetheless, their frequency
remains very small, reaching a maximum of only
$0.8\%$ for normalized dynamical distances of
at least $0.3$. Each point is averaged over 1,000 realizations.}
\end{figure}

\begin{figure}[h]
\includegraphics[width=0.65\textwidth]{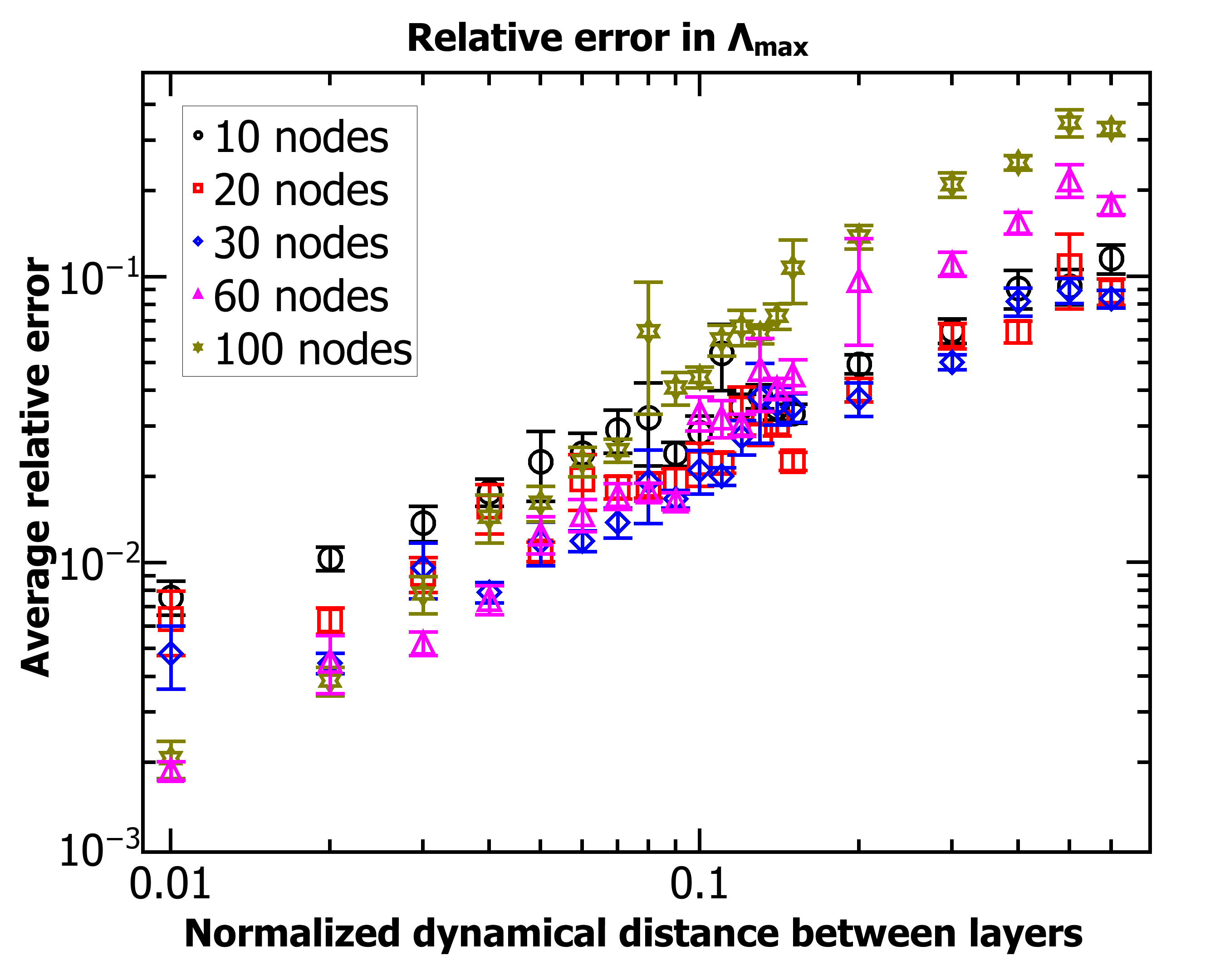}
\caption{{\bf The relative error in the largest Lyapunov
exponent in preferential-attachment networks has no clear
system-size dependence}.
The values of the average error in estimating the maximum
Lyapunov exponent remain similar to those obtained for random
networks, but without presenting a clear trend with network
size. Also, the variability of the error is appreciably larger
than it is for random networks. Each point is averaged over
1,000 realizations.}
\end{figure}